\documentclass[11pt,a4paper]{article}
\usepackage{jcappub}
\usepackage{varioref,exscale,latexsym,amsmath,amssymb}
\usepackage{graphicx,epstopdf}
\usepackage{comment}
\usepackage[usenames,dvipsnames]{xcolor}
\usepackage{epsfig}
\usepackage{graphicx}
\usepackage{dcolumn}
\usepackage{bm}
\usepackage{simplewick}

\def\beq{\begin{equation}}

\def\eeq{\end{equation}}

\def\beqa{\begin{eqnarray}}

\def\eeqa{\end{eqnarray}}

\begin{document}

\title{Inflationary magnetogenesis and non-local actions: The conformal anomaly}

\author{Basem Kamal El-Menoufi}

\affiliation{Department of Physics,
University of Massachusetts\\
Amherst, MA  01003, USA}

\emailAdd{bmahmoud@physics.umass.edu}

\abstract{
We discuss the possibility of successful magnetogenesis during inflation by employing the one-loop effective action of massless QED. The action is strictly non-local and results from the long distance fluctuations of massless charged particles present at the inflationary scale. Most importantly, it encodes the conformal anomaly of QED which is crucial to avoid the vacuum preservation in classical electromagnetism. In particular, we find a blue spectrum for the magnetic field with spectral index $n_B \simeq 2 - \alpha_{\text{e}}$ where $\alpha_{\text{e}}$ depends on both the number of e-folds during inflation as well as the coefficient of the one-loop beta function. In particular, the sign of the beta function has important bearing on the final result. A low reheating temperature is required for the present day magnetic field to be consistent with the lower bound inferred on the field in the intergalactic medium.}

\maketitle
\flushbottom

\section{Introduction}

Magnetic fields of various strengths have been observed at several length scales in our Universe. For instance, in galaxies they are of the order of few $\mu$G. While the origin of galactic fields remains mysterious, it is widely accepted that a seed field which predates structure formation is required to produce the observed fields today. There exist several dynamo mechanisms able to amplify a relatively weak field to the currently observed field strengths \cite{Brandenburg}. A conservative estimate demands a field of strength $10^{-23} \, \text{G}$ at the $1 \, \text{Mpc}$ scale to appropriately seed the galactic dynamo.

On the other hand, magnetic fields also exist in the intergalactic medium (IGM) where recent bounds have been inferred in \cite{Neronov,Taylor,Vovk} from the lack of observation of GeV electromagnetic cascades initiated by TeV gamma rays in the IGM. These fields are especially interesting from a cosmological standpoint since it is unlikely that they are due to some astrophysical mechanism \cite{Durrer}. These measurements have thus re-opened the door to further investigate a primordial origin of cosmic magnetic fields. Although astrophysical processes could generate the required seed for the galactic dynamo, a primordial origin remains an attractive possibility as well in this case. Having the ability to amplify quantum fluctuations, the inflationary epoch offers the perfect setting to establish magnetogenesis in the early Universe. Moreover, understanding inflationary magnetogenesis could help better constrain the landscape of inflationary models if future experiments confirmed the primordial origin of cosmic magnetic fields.     

One obstacle to achieving inflationary magnetogenesis is the conformal invariance of classical electromagnetism \cite{Turner}. In a spatially flat Universe, this implies that the conformal vacuum is preserved and magnetic fields can not be amplified \cite{Demozzi}. Thus the starting point in model building is the breakage of conformal invariance. Ratra \cite{Ratra} studied a model with the gauge-invariant Lagrangian $-\frac{1}{4}I^2 F_{\mu\nu} F^{\mu\nu}$ where $I$ is a function of conformal time. For instance, a coupling of the gauge field to the inflaton during slow-roll would give rise to such scenario \cite{Martin}. Another proposed mechanism is the axion model \cite{Carroll} where a psuedoscalar inflaton is coupled to $F_{\mu\nu} \tilde{F}^{\mu\nu}$. Parity violation in particular has the advantage of producing {\em maximally helical} fields \cite{Anber} whose coherence scale grows much faster than non-helical fields during cosmic evolution. A recently proposed model \cite{Caprini} is a hybrid of the previous ones where a time dependent function appears in front of both the parity-preserving and partity-violating invariants. A UV realization of the latter model was proposed in the context of $\mathcal{N}=1$ four-dimensional supergravity.

On the other hand, the breakdown of conformal symmetry takes place naturally due to vacuum fluctuations. Although less appreciated, one important aspect of anomalies is their {\em infrared} origin. It is precisely the low energy portion of quantum loops of massless particles that breaks the classical symmetry. In particular, this implies that any new physics that might appear in the UV would not alter the anomaly structure. By using dispersive techniques, this piece of physics was originally emphasized in \cite{Zakharov,Frishman} in the context of the axial anomaly and later by \cite{Horejsi2} for the conformal anomaly.

Recently in \cite{Basem3}, the infrared physics of the conformal anomaly was developed further by constructing the effective action of massless QED. The anomaly could be elegantly reproduced from the effective action that results from integrating out the massless charged particle. Being produced by long-distance fluctuations, the renormalized effective action is non-local in position space. Non-local field theories have just started to be explored with various applications especially in cosmology \cite{Basem1,Espriu,Cabrer,Woodard1,Deser,Calmet,Conroy,Maggiore}, which is surely an incomplete list. However, the construction of non-local actions over curved spaces is far from being trivial. For the QED case, this has been systematically carried out in \cite{Basem2}. 

To our knowledge, Dolgov \cite{Dolgov} made the first attempt to employ the conformal anomaly to derive inflationary magnetogenesis\footnote{See \cite{Agullo} for a similar treatment using purely gravitational anomalies.}. Only knowledge of the {\em local} anomalous operator was used in \cite{Dolgov} with the strength of the field required to seed the galactic dynamo being highly dependent on the sign of the one loop beta function of an $SU(N)$ gauge theory\footnote{Note that the author in \cite{Dolgov} uses the beta function of the full $SU(N)$ gauge theory as the coefficient of the anomalous operator. We argue that this is inaccurate since one must formally use the electric charge beta function.}. It is also worth mentioning that mechanisms including {\em local} gravitational-electromagnetic couplings were explored by many authors, see for example \cite{Kunze,Campanelli,Mazzitelli}. These couplings would naturally arise if massive fields heavier than the inflationary scale are present. Another mechanism was discussed in \cite{Prokopec} where a radiatively induced photon mass during inflation is used to generate magnetic fields.

In this paper, we are concerned with the non-local action that generates the QED trace anomaly. If the Standard Model (SM) electroweak symmetry is unbroken during inflation, then all charged fermions are massless. In particular, integrating out the latter yields a non-local action which encodes the conformal anomaly\footnote{To be precise, a photon is not an active degree of freedom in the unborken phase. Our presentation is exploratory in this regard and more comments appear in the concluding remarks.}. We present a thorough analysis to investigate the viability of magnetogenesis during inflation using the QED trace anomaly as the driving mechanism. Although the action is very complicated, we will see that the anomalous portion is rather simple to handle during inflation assuming an exact de-Sitter phase. In particular, the constancy of the scalar curvature enables the action to be written in a form similar to the models previously described which simplifies the analysis greatly. Despite this similarity, it is important to realize that the mechanism discussed here does not require any physics beyond the SM. The action is parameter-free and thus we need not worry about any possible constraints usually discussed in the model building literature. We find a rather blue spectrum at the end of inflation given that the QED beta function is positive. The evolution of the initial conditions till the present day is carried out via two pathways. We first evolve the magnetic field based on the simple requirement of flux conservation. The reheating temperature has to be relatively low to satisfy the lower bound on the IGM field reported in \cite{Neronov,Taylor,Vovk}. Second, we summarize the main features of the magneto-hydrodynamic (MHD) evolution \cite{Durrer,Jedamzik} and argue that the simple evolution is largely accurate with our initial conditions.

The plan of the paper is as follows. We describe in some detail the non-local effective action in section \ref{sect2}. Then in section \ref{sect3} we describe how to cast the non-local action in a simple form. The theory is canonically quantized and approximate solutions for the mode functions of the gauge field are found in section \ref{sect4}. In section \ref{sect5} the properties of the magnetic field at the end of inflation are determined. In section \ref{sect6} the evolution of the initial conditions are carefully carried out. In section \ref{sect7}, we test whether the present day properties of the magnetic field are consistent with the lower bound in \cite{Neronov,Taylor,Vovk}. We conclude and discuss future directions in section \ref{conc}.     

\section{The non-local action}\label{sect2}

The effective action is an extremely useful object in field theory, in particular, it embodies all the effects of quantum fluctuations. By construction it is the generating functional of one-particle irreducible (1PI) correlation functions. Its prominent use is when the problem involves classical background fields and one aims to study the effect of quantum loops in a semiclassical context. In particular, its importance in gravitational physics can not be overestimated \cite{Buchbinder,Birrell,parkertoms}. Formally, one computes the effective action by integrating out a field from the path integral of the theory. If the field is heavy, the result is a local effective Lagrangian built from the light degrees of freedom organized in a derivative (energy) expansion and the cut-off of the effective theory is the mass of the heavy field. On the other hand, loops of massless fields leads to non-analyticity in momentum space or equivalently non-locality in position space \cite{Donoghue1,Donoghue2}. These effects strictly arise from the infrared fluctuations of massless particles and the resulting effective Lagrangian is non-local. There has been a consistent effort to understand the construction, properties and phenomenology of non-local Lagrangians.

Anomalies in field theory remains to date an active area of research due to their wide array of applications. The common lore in the literature is that anomalies are understood through the UV properties of Feynman diagrams. Using different approaches, several authors pointed out that it is the low-energy portion of quantum loops that give rise to anomalies \cite{Zakharov,Horejsi2,Basem3}. In the gravity sector, the seminal work of Deser, Isham and Duff \cite{Isham} was the first attempt to reproduce gravitational anomalies from a non-local action. On the gauge theory side and in the context of massless QED, both the non-local action and the associated energy-momentum tensor (e.m.t) were constructed in \cite{Basem3} with the initial results displayed for flat space. Subsequently, these results were carried over to curved space employing a technique referred to as {\em non-linear completion} \cite{Basem2}. The latter shares similar features with the Covariant Perturbation Theory formalism developed by Barvinsky, Vilkovisky and collaborators \cite{Barvinsky1,Barvinsky2,Barvinsky3}. 
  
Here we only quote the main results and refer the interested reader to \cite{Basem2} for more details. The classical theory under consideration is QED coupled to either charged scalars or fermions. For instance, in case of a charged scalar the classical action reads
\begin{align}\label{scalaraction}
S = S_{EM} + \int d^4x \sqrt{g} \big[g^{\mu\nu}(D_{\mu}\phi)^{\star}(D_{\nu}\phi) - \xi \phi^{\star}\phi R \big]
\end{align}
where $S_{EM}$ is the standard maxwell action, $D_\mu = \partial_\mu + ie_0 A_\mu$ is the gauge-covariant derivative and $e_0$ is the bare electric charge. For $\xi=1/6$, the action is indeed invariant under local Weyl transformations
\begin{align}\label{weyltrans}
g_{\mu\nu} \rightarrow e^{2\sigma(x)} g_{\mu\nu}, \quad \phi \rightarrow e^{-\sigma(x)} \phi, \quad A_{\mu} \rightarrow A_{\mu} \ \ .
\end{align}
Conformal invariance is manifest in the tracelessness of the classical e.m.t. After integrating out the massless charged field, one ends up with a variety of terms that exhibits different behavior under conformal and scale transformations. It was shown in \cite{Basem2} that the piece that ultimately generates the anomaly is given by 
\begin{align}\label{anomactionQL}
\Gamma_{anom.}[g,A] = S_{EM} - \frac{b_i e^2}{12} \int d^4x \, \sqrt{g} F_{\mu\nu} F^{\mu\nu} \frac{1}{\nabla^2} R \ \ .
\end{align}
Here $b_i$ is the leading coefficient of the electric charge beta function
\begin{align}
b_s = \frac{1}{48\pi^2}, \quad b_f = \frac{1}{12\pi^2}
\end{align}
and $\nabla^2 = g^{\mu\nu} \nabla_\mu \nabla_\nu$ is the covariant d' Alembertian. We also include the Maxwell action for consistency and later usage. To see how the anomaly arises, we employ an infinitesimal conformal transformation given by
\begin{align}\label{conftrans}
\delta_\sigma g_{\mu\nu} = 2 \sigma g_{\mu\nu}, \quad \delta_\sigma R = 6 \nabla^2 \sigma - 2 \sigma R \ \ .
\end{align}
A generic action transforms as follows
\begin{align}
\delta_{\sigma} S = - \int d^4x \sqrt{g} \sigma T_\mu^{~\mu} \ \ .
\end{align}
and thus transforming the action in eq. (\ref{anomactionQL}) immediately yields the correct trace relation\footnote{The details of these steps are explained clearly in \cite{Basem2}.}
\begin{align}
T_\mu^{~\mu} = \frac{b_i}{2} F_{\mu\nu} F^{\mu\nu}
\end{align}
Written in this form, we say that the action in eq. (\ref{anomactionQL}) is {\em quasi-local}. In purely non-local form, we have
\begin{align}\label{anomactionNL}
\Gamma_{anom.}[A] = S_{EM} - \frac{b_i e^2}{12} \int d^4x \, d^4y \, \sqrt{g(x)} \sqrt{g(y)} (F_{\mu\nu} F^{\mu\nu})_x G(x,y) R_y 
\end{align}
where the propagator satisfies
\begin{align}\label{propeqn}
\nabla^2_x \, G(x,y) = \frac{\delta^{(4)}(x-y)}{(\sqrt{g(x)}\sqrt{g(y)})^{1/2}} \ \ .
\end{align}

\section{The set-up}\label{sect3}

Many models of magnetogenesis start with the following Lagrangian \cite{Ratra,Demozzi,Carroll,Caprini}
\begin{align}\label{action}
\mathcal{L} = I^2(\tau)\left(-\frac{1}{4} F_{\mu\nu} F^{\mu\nu} \right)
\end{align}
where $I(\tau)$ is some specified function that contains the parameters of the model and indices are raised and lowered using the flat metric. Inspection of eq. (\ref{anomactionNL}) shows that we can cast the action in the form of eq. (\ref{action}) since the scalar curvature is constant during an exact de-Sitter phase. This however requires knowledge of the propagator on a de-Sitter background which fortunately could be obtained in closed form. We show in this section how to manipulate eq. (\ref{anomactionNL}) to identify the function $I^2(\tau)$.
 
We work in the cosmological slice of de-Sitter and write the metric in conformal coordinates
\begin{align}
ds^2 = a^2(\tau) \left(d\tau^2 - d\vec{x} \cdot d\vec{x} \right), \quad a(\tau) = (-H\tau)^{-1}, \quad -\infty < \tau < 0 \ \ .
\end{align} 
We start by solving for the propagator in eq. (\ref{propeqn}) where we impose the usual retarded boundary conditions\footnote{It has been shown in \cite{Basem1} that using the in-in formalism yields a causal prescription for the non-local functions.}. In the above metric, eq. (\ref{propeqn}) becomes
\begin{align}
\frac{1}{a^2(\tau)} \left(\partial^2 + \frac{2 a^\prime}{a} \partial_\tau \right) G(x,y) = \frac{\delta^{(4)}(x-y)}{a^2(\tau) a^2(\tau^\prime)}
\end{align}
where $\tau = x^0$ while $\tau^\prime = y^0$. It suffices to determine the inverse of the operator appearing in brackets on the lhs. With flat spatial slices, the propagator can be expanded as a Fourier integral
\begin{align}
G(x,y) = \int\frac{d^3\vec{k}}{(2\pi)^3}\, G(\tau,\tau^\prime; k) e^{i\vec{k}\cdot(\vec{x}-\vec{y})} \ \ .
\end{align}
Hence, the function $G(\tau,\tau^\prime; k)$ satisfies the equation
\begin{align}
\left(\frac{d^2}{d\tau^2} + k^2 - \frac{2}{\tau} \frac{d}{d\tau} \right) G(\tau,\tau^\prime; k) = \frac{\delta(\tau - \tau^\prime)}{a^2(\tau^\prime)} \ \ .
\end{align}
The retarded propagator of the operator in brackets is well known \cite{Caprini}. Hence, our function reads
\begin{align}
G(\tau,\tau^\prime; k) = \frac{H^2}{k^3} \left((1 + k^2 \tau \tau^\prime) \sin k(\tau-\tau^\prime) + k (\tau^\prime - \tau) \cos k (\tau-\tau^\prime) \right) \Theta(\tau - \tau^\prime)\ \ .
\end{align}
We finally plug everything back in the action eq. (\ref{anomactionNL}) and notice that the $d^3\vec{y}$-integral is trivial since the scalar curvature is constant. The integral yields a delta function $\delta^{(3)}(k)$ and so we must first expand the propagator around $\vec{k} = 0$ to find
\begin{align}
\Gamma_{anom.}[A] = S_{EM} - \frac{R b_i e^2}{36 H^2}  \int d^4 x \, \sqrt{g} F^2 \int \frac{d\tau^\prime}{\tau^{\prime \, 4}} \, (\tau^3 - \tau^{\prime \, 3}) \Theta(\tau - \tau^\prime) \ \ .
\end{align}
It is gratifying to see that the answer is completely well-behaved. Now $R = 12H^2$ and thus we can now identify the time-dependent function
\begin{align}
I^2(\tau) = 1 + \frac{4 b_i e^2}{3} \int \frac{d\tau^\prime}{\tau^{\prime \, 4}} \, (\tau^3 - \tau^{\prime \, 3}) \Theta(\tau - \tau^\prime) \ \ .
\end{align}
The second piece in the bracket leads to a logarithmic divergence\footnote{This is not surprising since the long-time behavior of the non-local functions corresponds to the far infrared. When integrated against a source, the long-time behavior of the result might become singular.}. However, this causes no trouble since it is always plausible to cut off the integral at an early time $\tau_0$ corresponding to the beginning of inflation. The effect of this arbitrary parameter on the physical observables we consider is thoroughly discussed in subsequent sections. Finally we obtain
\begin{align}\label{ioftau}
I^2(\tau) = 1 + \frac{4 b_i e^2}{3} \left[ \frac{1}{3} \left(\frac{\tau}{\tau_0}\right)^3 -\frac13 + \ln \left(\frac{\tau_0}{\tau}\right)\right] \ \ .
\end{align}
It is desirable to pause at this stage and comment on some issues regularly discussed in the model building literature. The first aspect concerns whether the theory is strongly coupled \cite{Demozzi}. We easily see that $I^2(\tau) \geq 1$ during inflation and hence we are definitely in a weak coupling regime. This is guaranteed with a positive definite beta function. But let us now imagine that $I^2(\tau)$ was in fact less than unity due to a negative beta function. Even in this hypothetical situation, no problem arises in our case. The effective action is the result of integrating out the massless charged particles and thus, formally, the latter can not appear as external states in the theory. We argue that unlike magnetogenesis models the issue of strong coupling does not posit a concern all together. Along the same lines, it was shown in \cite{Barnaby,Namba} that a serious challenge to magnetogenesis models emerges if the time-dependent function is the result of coupling the gauge field to the rolling inflaton. In this case the amplified gauge field couples to the inflaton perturbations leading to observable nongaussianities which provides an extra constraint on the parameters of any such model. Such constraints do not apply in our case as the action is parameter-free and relies only on the existence of massless charged particles during inflation. Nevertheless, the non-local coupling in eq. (\ref{anomactionQL}) inevitably contribute to the curvature perturbation\footnote{For example, see the construction in \cite{Dimo}.}. This is one exciting direction that we leave for the future.

Let us now include the effect of multiple particles in the loop and define the following constant
\begin{align}
\beta \equiv \frac{4}{3} \left( \sum_f  g_f b_f Q_f^2 + \sum_s  g_s b_s Q_s^2 \right) \ \ .
\end{align}
where $g_f(g_s)$ is the number of fermionic (scalar) internal degrees of freedom and $Q_f(Q_s)$ is the electric charge of each species. We can now rewrite eq. (\ref{ioftau}) as
\begin{align}\label{ioftau2}
I^2(\tau) = 1 + \beta \left[ \frac{1}{3} \left(\frac{\tau}{\tau_0}\right)^3 -\frac13 + \ln \left(\frac{\tau_0}{\tau}\right)\right] \ \ .
\end{align}
To get an idea about the range of values $\beta$ can take, let us restrict to the charged fermions in the Standard Model and find
\begin{align}
\beta_{SM} =  \frac{4}{3} \sum_l \frac{Q^2_l}{12\pi^2} + 4 \sum_q \frac{Q^2_q}{12\pi^2} 
\end{align}
where $l$ and $q$ refer to leptons and quarks respectively. We now use the one-loop beta function to run the electric charge from the weak scale up to the energy scale of inflation. Hence
\begin{align}
\frac{1}{e^2(E_{\text{inf}})} = \frac{1}{e^2(M_Z)} - \frac{4}{3\pi^2} \ln\left(\frac{E_{\text{inf}}}{M_Z}\right)
\end{align}
where $M_Z$ is the Z-boson mass. Using input from \cite{Agashe} and taking the energy scale of inflation to be above the electroweak scale yields
\begin{align}\label{betasm}
\beta_{SM} \simeq 10^{-2} \div 10^{-3} \ \ .
\end{align}

\section{Canonical Quantization}\label{sect4}

In this section we perform the quantization procedure and find approximate solutions to the mode functions. It is straightforward to derive the equations of motion from the action in eq. (\ref{action}) 
\begin{align}\label{eombasic}
\partial^\mu \left(\eta^{\beta\nu} I^2 \, F_{\mu\nu} \right) = 0
\end{align}
where $\partial^\mu = \eta^{\mu\alpha} \partial_\alpha$ and we kept a flat metric inside the brackets manifest so that no confusion arises. In the absence of currents, we employ Coulomb gauge $\partial^i A_i = 0$ that forces $A_0 = 0$ and hence eq. (\ref{eombasic}) becomes
\begin{align}\label{eomgauge}
\left(\partial^2 + \frac{2 I^\prime}{I} \partial_\tau\right) A_i = 0, \quad \partial^i  A_i = 0 \ \ .
\end{align}
The quantization of the gauge field proceeds as usual via the canonical formalism
\begin{align}\label{gaugeoperator}
\hat{A}_i(x) = \sum_{\sigma=1,2} \int \frac{d^3\bold{k}}{(2\pi)^{3/2}} \epsilon_i(\bold{k},\sigma) a(\bold{k},\sigma) A(k,\eta) e^{i\bold{k}\cdot\bold{x}} + h.c.
\end{align}  
where $a(k,\eta)$ and $a^\dagger(k,\eta)$ are creation and annihilation operators satisfying 
\begin{align}
[ a(\bold{k},\sigma), a^\dagger(\bold{k}^\prime,\sigma^\prime) ] = \delta^{(3)} (\bold{k} - \bold{k}^\prime) \delta_{\sigma\sigma^\prime} \ \ .
\end{align}
Indeed the polarization tensors are transverse but notice here that they are covariantly normalized, in particular, they carry explicit time dependence \cite{Martin}
\begin{align}
\epsilon(\bold{k},\sigma) \cdot \epsilon(\bold{k},\sigma^\prime) = -\delta_{\sigma\sigma^\prime} \ \ .
\end{align}
Now we can define a canonically normalized mode function by
\begin{align}
\tilde{A}(\bold{k},\eta) = aI A(\bold{k},\eta) \ \ .
\end{align}
The reason the scale factor is inserted is to cancel the time dependence explicit in the polarization tensors. Now applying eq. (\ref{eomgauge}), we find
\begin{align}\label{modefunc}
\left(\partial_\tau^2 + k^2 - \frac{I^{\prime\prime}}{I} \right) \tilde{A}(k,\tau) = 0 \ \ .
\end{align}
The power spectrum is readily found from the two-point function which reads
\begin{align}
\langle 0 | \hat{A}^\mu (\tau,\vec{x}) \hat{A}_\mu (\tau,\vec{y}) | 0 \rangle = -\frac{2}{ a^2 I^2} \int \frac{d^3k}{(2\pi)^{3/2}} \, \tilde{A}(k,\tau) \tilde{A}^*(k,\tau) e^{i \vec{k} \cdot (\vec{x} -\vec{y})} \ \ .
\end{align} 
From the coincidence limit, we determine the power spectrum
\begin{align}
\mathcal{P}_A(k,\tau) = \sqrt{\frac{k^3 |\tilde{A}(k,\tau)|^2}{2\pi^2 a^2 I^2}} \ \ .
\end{align}

\subsection{Solving for the mode functions}

Solving eq. (\ref{modefunc}) exactly is not possible due to the non-trivial nature of $I(\tau)$. Nevertheless, all what we really need is an approximate solution at the end of inflation which is sufficient to determine the power spectrum as well as the amplitude of the magnetic field and its correlation length, i.e. the initial conditions. First of all, we easily find
\begin{align}\label{iprime}
\frac{I^{\prime\prime}}{I} = \frac{\beta}{2 I^2} \frac{1}{\tau^2} \left[\left(1 + \frac{2\tau^3}{\tau_0^3}\right) - \frac{\beta}{2I^2} \left(1 - \frac{\tau^3}{\tau_0^3} \right)^2 \right] \ \ .
\end{align}
At the onset of inflation $(\tau \sim \tau_0)$ all modes of cosmological interest are inside the horizon, i.e. $k |\tau| \gg 1$, and hence the modes reside in the Bunch-Davies vacuum. The positive energy solution to eq. (\ref{modefunc}) reads
\begin{align}\label{BD}
\tilde{A}(\tau,k) \simeq \frac{1}{\sqrt{2k}} e^{-i k (\tau - \tau_i)}, \quad \tau \to \tau_0
\end{align}
where $\tau_i$ is arbitrary and will later be chosen for convenience. As the size of the horizon decreases, the modes start to leave their vacuum state and get amplified. When a mode approaches horizon exit, we can approximate
\begin{align}\label{iprimeapp}
\frac{I^{\prime\prime}}{I} \simeq \frac{\beta}{2 I^2} \frac{1}{\tau^2} \left[1 - \frac{\beta}{2I^2} \right]
\end{align}
valid because $(\tau/\tau_0)^3 \ll 1$ at this stage. We can further process the above expression if we notice that
\begin{align}
I^2 \simeq 1 + \beta \, N(\tau)
\end{align}
where $N(\tau)$ is the number of e-folds since the beginning of inflation. Thus the second term in the brackets in eq. (\ref{iprimeapp}) is much smaller than unity and could be dropped. This turns eq. (\ref{modefunc}) into a rather simple form
\begin{align}\label{modehc}
\left(\partial^2_\tau + k^2 - \frac{\alpha(\tau)}{\tau^2} \right) \tilde{A}(k,\tau) = 0, \quad |\tau| \lesssim |\tau_k| = 1/k
\end{align}
where we defined
\begin{align}\label{varalpha}
\alpha(\tau) \equiv \frac{\beta}{2[1 + \beta\, N(\tau)]} \ \ .
\end{align}
Eq. (\ref{modehc}) is readily solved with Bessel functions if $\alpha$ was constant. Can we treat $\alpha$ as a constant? It is reasonable to adopt this approximation as the rate of change of the last term in eq. (\ref{modehc}) is controlled by\footnote{One can easily check this statement by taking a time derivative of the aformentioned term and using the fact that after a few e-folds $\alpha$ becomes negligible {\em compared to unity}.} $1/\tau^2$. Hence
\begin{align}
\tilde{A}(k,\tau) \simeq \frac{1}{\sqrt{k}} \left[ c_1 \, (-k\tau)^{1/2} {\rm J}_\nu (-k\tau) + c_2 \, (-k\tau)^{1/2} {\rm J}_{-\nu} (-k\tau) \right], \quad \nu = \frac12 \sqrt{1 + 4 \alpha}
\end{align} 
where $c_1$ and $c_2$ are constants to be determined. Notice here that the order of the Bessel functions is treated as {\em time-dependent}. We match the solutions and their first derivative at the time of horizon crossing, i.e. $\tau_k = - 1/k$, onto the free solutions in eq. (\ref{BD}). Fixing $\tau_i = \tau_k$ we find
\begin{align}\label{coeff}
\vec{c} = \hat{\gamma}^{-1} \vec{r}
\end{align}
where $\vec{c}^{~T} = (c_1, c_2)$ and $\vec{r}^{~T} = (1/\sqrt{2}, i/\sqrt{2})$ while the matrix $\hat{\gamma}$ reads
\begin{align}
\nonumber
\gamma_{11} &= {\rm J}_{\nu_k}(1) & \gamma_{21} &= \frac12 {\rm J}_{\nu_k}(1) + {\rm J}^\prime_{\nu_k}(1) \\
\gamma_{12} &= {\rm J}_{-\nu_k}(1)& \gamma_{22} &= \frac12 {\rm J}_{-\nu_k}(1) + {\rm J}^\prime_{-\nu_k}(1) \ \ .
\end{align}
Here, $\nu_k = (1+4\alpha(\tau_k))/2$ is determined by the number of e-folds until a certain mode crosses the horizon. One can easily check that the coefficients $c_1$ and $c_2$ are $\mathcal{O}(1)$ complex numbers and thus will not change our final results in any significant manner. Now at the end of inflation $(\tau \to \tau_e)$, all modes of cosmological interest are way outside the horizon implying $-k\tau \ll 1$ and thus we can approximate the Bessel functions and turn the result into a power law. The solution multiplying $c_1$ contribute negligibly for the considered modes and hence the mode functions take the rather simple form
\begin{align}\label{modefuncend}
\tilde{A}(k,\tau_e) \simeq \frac{\bar{c}_{\tiny 2}}{\sqrt{k}} (-k\tau_e)^{(1- \sqrt{1+4\alpha_e})/2}, \quad \bar{c}_2 = \frac{2^{1/2\sqrt{1+4\alpha_e}}}{\Gamma(1-1/2\sqrt{1+4\alpha_e})}\, c_2 
\end{align}
where $\alpha_{\text{e}} = \alpha(\tau_{\text{e}})$ is given in terms of the total number of e-folds during inflation. Indeed $\bar{c}_2$ is an $\mathcal{O}(1)$ number as well.

\section{The magnetic field at the end of inflation}\label{sect5}

Our task in this section is to determine the properties of the magnetic field at the end of inflation: the amplitude of the field, its coherence scale and the spectral index. These initial conditions will be subsequently evolved to the present time. We start from the covariant definition of the magnetic field in curved space \cite{Martin}
\begin{align}
B_\mu = \frac{1}{2} \epsilon_{\mu\nu\alpha\beta} \, u^\beta F^{\nu\alpha}
\end{align}
where $u^\mu$ is the $4$-velocity vector field tangent to an observer's worldline and $\epsilon_{\mu\nu\alpha\beta}$ is the totally antisymmetric tensor, i.e. $\epsilon_{0123} = \sqrt{g}$. For a comoving observer $u^\mu = (1/a,\bold{0})$ and
\begin{align}\label{bfield}
B_i = \frac{1}{a} \epsilon_{ijk} \partial_j A_k \ \ . 
\end{align}
It is now straightforward to find the square of the magnetic field power spectrum from the two-point function
\begin{align}\label{magspectrum}
\mathcal{P}_B(k,\tau) = \sqrt{\frac{k^5 |\tilde{A}(k,\tau)|^2}{\pi^2 a^4 I^2}}  \ \ .
\end{align}
Notice the extra power of the scale factor in the denominator relative to the gauge field power spectrum. Plugging in the solution in eq. (\ref{modefuncend}), we can easily read off the spectral index
\begin{align}
\nonumber
n_B &= 2 + \frac{1}{2}\left( 1 - \sqrt{1 + 4\alpha_{\text{e}}} \right)  \\
&\simeq 2 - \alpha_{\text{e}}
\end{align} 
valid since $\alpha_{\text{e}} < 1$. A precise knowledge of the spectral index is crucial to determine the strength of the magnetic field at the present epoch and thus one should investigate at this stage the exact size of $\alpha_e$. The total number of e-folds strongly depends on the dynamics of inflation \cite{Remmen} so we are going to fix $N=60$ since, as evident from eq. (\ref{varalpha}), lowering the number of e-folds yields a larger $\alpha_{\text{e}}$. Moreover, to obtain a {\em best value} we will imagine dialing up the number of particles in the loop such that the spectral index asymptotes to\footnote{Working instead with $\beta_{SM}$ does not alter the spectral index significantly.}
\begin{align}\label{alphae}
n_B \simeq 1.991 \ \ .
\end{align}
It is rather important to pause at this stage and notice that {\em reversing} the sign of the beta function would change the whole picture. If $\beta$ is small but negative one would be able to achieve a noticeably larger $\alpha_{\text{e}}$ and in turn a spectrum which is less blue. In fact, one could even obtain a (nearly) scale-invariant spectrum by adjusting the number of particles, a result that might be enough to generate the present day IGM field as well as to ignite the galactic dynamo \cite{Durrer}. This somewhat echoes the observation made in \cite{Dolgov} and we reserve considering this possibility to a future publication. 

The second quantity of interest is the average strength of the magnetic field which reads
\begin{align}\label{bfieldfourier}
\text{B}^2(\tau_e) = \left(\pi I_e a_e^2 \right)^{-2} \int_{k_{\min}}^{k_{\max}} dk \, k^4 \, |\tilde{A}(k,\tau_e)|^2 
\end{align}
where $k_{\min}$($k_{\max}$) is an IR(UV) cut-off. The value of $k_{\max}$ is naturally dictated by the size of the horizon at the end of inflation, namely $k_{\max} = H a_e$ corresponding to the last mode that crossed the horizon and felt the amplification. On the other hand, strictly speaking $k_{\min}$ should be determined by the size of the horizon {\em today} but for simplicity we are going instead to take $k_{\min} = (|\tau_0|)^{-1}$. This choice does not alter the result as we show next. The above integral could readily be performed and yields 
\begin{align}\label{Bfieldend}
\text{B}^2(\tau_e) = \frac{\mathcal{O}(1)}{(4-2\alpha_e)\pi^2 I^2_e}\, H^4 \ \ .
\end{align}
Indeed, the coefficient $\bar{c}_2$ depends implicitly on the wavenumber and should have been included in the integral but this complicates the analysis without gaining any insight. The lower limit of the integral contributes negligibly to the amplitude and thus the precise choice of the IR cut-off does not affect the result, which is a manifestation of the blue spectrum.

Finally we need the {\em comoving} coherence scale of the magnetic field at the end of inflation. As we describe in the next section, the value of the present day magnetic field is determined by the evolution of the coherence scale. It is defined as \cite{Durrer,Jedamzik}
\begin{align}
\lambda_B(\tau_e) = 2\pi \, \frac{\int dk \, k^{-1} \, B^2(k,\tau_e)}{\int dk \, B^2(k,\tau_e)}
\end{align} 
where $\bold{B}(k,\tau_e)$ is the Fourier decomposition of the magnetic field. Performing the integrals, we find
\begin{align}\label{cohscale}
\lambda_B(\tau_e) = \mathcal{O}(1) \frac{(4-2\alpha_e)}{(3-2\alpha_e)} \frac{2\pi}{Ha_e} \ \ .
\end{align}

\section{The current magnetic field}\label{sect6}

The results of the previous section provides the initial conditions for the subsequent evolution of the magnetic field. As is well known \cite{Durrer,Jedamzik}, to trace the exact evolution of the magnetic field is quite complicated. The conventional treatment is to assume that the magnetic field {\em freezes} in the cosmic plasma quickly after inflation ends. This is because the electric conductivity of the plasma becomes effectively infinite leading the gauge field to become almost static after inflation \cite{Martin}. Inspection of the power spectrum eq. (\ref{magspectrum}) shows that the magnetic field is simply diluted by the scale factor squared which is nothing but the requirement of magnetic flux conservation.  

In this simple picture it suffices to know the ratio $(a_0/a_{\text{end}})$ where $a_0$ is the scale factor today while $a_{\text{end}}$ is that at the end of inflation. This ratio precisely depends on three {\em independent} parameters: the energy scale of inflation, the reheating temperature and the equation of state parameter during reheating \cite{Ringeval}. It reads
\begin{align}\label{scaleratio}
\frac{a_{\text{end}}}{a_0} = \mathcal{R} \left(\Omega^0_{\text{rad}} \frac{3 H_0^2}{M_P^2} \right)^{1/4} \left(\frac{\rho_{\text{end}}}{M_P^4} \right)^{-1/2}, \quad \rho_{\text{end}} = 3 H^2 M_P^2 \ \ .
\end{align}
The parameter $\mathcal{R}$ is a function of the three variables $(w_{\text{reh}},T_{\text{reh}},\rho_{\text{end}})$ and it determines the amplitude and coherence scale of the present day magnetic field. Notwithstanding, its precise form is not important for our analysis but rather the range of values it could take. A {\em model-independent} estimate for the latter was carried out in \cite{Martin}
\begin{align}\label{Rvalues}
\frac{1}{4} \ln \left(\rho_{\text{nuc}}/M_P^4 \right) < \ln \mathcal{R} < -\frac{1}{12} \ln \left(\rho_{\text{nuc}}/ M_P^4 \right) + \frac{1}{3} \ln \left(\rho_{\text{end}}/ M_P^4 \right)
\end{align}
where $\rho_{\text{nuc}}$ is the radiation energy density at nucleosynthesis. Both the upper and lower bounds assume the lowest possible reheating temperature. The lower bound assumes $w_{\text{reh}} = -1/3$ while the upper bound assumes\footnote{This equation of state is realizable in models based on the {\em quintessential inflation} scenario \cite{Peebles} where, after inflation, the kinetic energy of the inflaton dominates the enrgy density.} $w_{\text{reh}} = 1$. It is clear from eq. (\ref{scaleratio}) that the larger $\mathcal{R}$ becomes the stronger the present day magnetic field would be. Yet, inspection of eq. (\ref{cohscale}) sjows that a larger $\mathcal{R}$ leads to a shorter coherence scale. We shall see in the next section how to obtain a lower bound on $\mathcal{R}$.

However, this simple picture of the evolution is inaccurate as was first pointed out by Banerjee and Jedamzik in \cite{Jedamzik}. The coupling of the magnetic field to the cosmic plasma results in non-linear energy {\em cascades} in Fourier space. In particular, the above estimate does not describe the physics at the coherence scale. To obtain a more robust prediction, one ideally has to evolve the non-linear magneto-hydrodynamical equations from the moment of genesis to the present day. It is needless to say that this is impossible to perform analytically. Fortunately, numerical simulations show that the gross features of the evolution is rather simple to understand \cite{Jedamzik,Durrer}.

The magnetic field evolves in three main stages depending on the initial conditions and the properties of the plasma. We briefly state the main features of each phase.
\begin{itemize}
\item Free turbulent decay: This phase is characterized by a large Reynolds number. The latter is given by \cite{Durrer}
\begin{align}\label{reynolds}
\text{R}_k(T) = \frac{v_k \, \lambda_k}{\lambda_{\text{mfp}}(T)}
\end{align}
where $v_k$ is the velocity of the fluid at some scale $\lambda_k$ and $\lambda_{\text{mfp}}$ is the {\em comoving} mean-free-path of the particles in the plasma. During this phase, the power spectrum at scales larger than $\lambda_B$ retains its original shape while at smaller scales the spectrum develops a universal slope \cite{Durrer} {\em irrespective} of the initial conditions. Overall, $\lambda_B$ grows while the amplitude decays.
\item Viscous phase: The system enters this phase once the mean-free-path of the least coupled particle becomes large enough that the Reynolds number becomes of order unity. The high viscosity suppresses plasma motions on scales up to the coherence scale. This leads the magnetic field to decouple from the plamsa. Overall, $\lambda_B$ stays constant and the magnetic field gets diluted only by expansion \cite{Jedamzik}.
\item Free streaming: Close to decoupling (e.g. neutrino decoupling), the mean-free-path grows beyond $\lambda_B$. Neutrinos, being too weakly coupled, do not provide {\em true} viscosity at this stage but rather contribute a friction term in the Euler equation \cite{Jedamzik}. The coefficient of the latter is inversely proportional to the mean-free-path and thus the turbulent phase is restored shortly before decoupling \cite{Durrer,Jedamzik}. Afterwards, the whole cycle is repeated but now with photons instead. 
\end{itemize}  
The magnetic field and the coherence scale evolve according to a power-law during turbulence and free-streaming \cite{Jedamzik}. The commencement/termination of each phase depends on the initial conditions and the properties of the plasma. Let us estimate the Reynolds number in eq. (\ref{reynolds}) right after inflation and for simplicity instantaneous reheating will be assumed. The {\em proper} mean-free-path in the plasma above the electroweak scale reads \cite{Durrer}
\begin{align}
l_{\text{mfp}} = \frac{22}{T} \ \ .
\end{align}
At the coherence scale $\lambda_B$, the velocity of the fluid is taken to be the Alfv\'{e}n speed \cite{Durrer} and thus
\begin{align}
\text{R}_{\lambda_B} \simeq \sqrt{H/M_P} \ll 1 \ \ .
\end{align}
Hence, the flow at the coherence scale is not turbulent with our initial conditions. As discussed in \cite{Durrer}, this condition is typical in inflationary magnetogenesis scenarios unless there exist a mechanism able to set the magnetic field in {\em equipartition} with the flow\footnote{The occurrence of parity violation is able to amplify the field to equipartition as shown in \cite{Caprini}.}. In particular, with our initial conditions the system starts in the viscous phase which means the magnetic field stays comovingly constant. For this reason it suffices to predict the present day amplitude and coherence scale based on flux conservation as we described above\footnote{It is possible that turbulence develops at a later stage in the evolution, e.g. at nuetrino decoupling. Yet, we do not consider such a possibility since it is unlikely that it affects our conclusion in any substantial manner.}.

\section{The lower bound on the IGM field}\label{sect7}

In this section, we employ the previous analysis to determine the properties of the present day magnetic field. In particular, we are concerned with satisfying the lower bound inferred on the IGM field which was given in \cite{Neronov,Taylor,Vovk}
\begin{align}\label{Bbound}
\text{B}_{\text{meas.}} \geq 6 \times 10^{-18} \sqrt{\frac{1\text{Mpc}}{\lambda_B}}\, \text{G}
\end{align} 
where account is taken of coherence scales shorter than $1 \, \text{Mpc}$. Notice that this is a combined bound on both the magnetic field and coherence scale, in particular, it does not constrain the spectral index. Using eqs. (\ref{Bfieldend}) and (\ref{scaleratio}) yields a present day magnetic field
\begin{align}\label{Bnownaive}
\text{B}_0 \simeq \frac{2 \times 10^{18}}{(1 + \beta N)^{1/2}}\, \Delta^2\, \text {G} 
\end{align}
and we defined the dimensionless quantity
\begin{align}
\Delta \equiv \frac{H}{1\text{GeV}} \frac{a_{\text{end}}}{a_0}\ \ .
\end{align} 
To obtain the best value, we obviously need to minimize the denominator in eq. (\ref{Bnownaive}) and thus we choose $N=60$ and $\beta = \beta_{SM}$. Using eq. (\ref{cohscale}) the bound in eq. (\ref{Bbound}) could be written as follows
\begin{align}\label{deltabound}
\Delta \gtrsim 10^{-34/3} \ \ .
\end{align}
Inspection of eq. (\ref{scaleratio}) reveals that the explicit dependence on the Hubble scale disappears from $\Delta$ all together. In fact, the above bound is readily turned into a lower bound on $\mathcal{R}$
\begin{align}\label{Rbound}
\ln \mathcal{R} \gtrsim 4 \ \ .
\end{align}
This is the main result of our analysis. Now one must inquire if this value for $\mathcal{R}$ is realizable. Assuming the highest possible scale of inflation, eq. (\ref{Rvalues}) leads to \cite{Martin}
\begin{align}
-47 \lesssim \ln \mathcal{R} \lesssim 10 \ \ .
\end{align}
We conclude that the QED trace anomaly is in principle capable of producing the IGM field although the reheating temperature must be very low. 

\section{Summary and conclusions}\label{conc}

Quantum loops of massless particles bring a unique feature to gravitational phenomena, i.e. non-locality. These effects have received recent interest in the literature especially in regard to cosmology. One open question of present day cosmology and astrophysics is the large-scale magnetic fields observed across our Universe. Such fields can not be produced by standard electromagnetism because conformal symmetry preserves the vacuum of the theory. As is well known, conformality is anomalously broken by loops of massless particles and precisely by the low energy portion of loops \cite{Horejsi2,Basem3}. It is important then to try achieving magnetogenesis using this basic field theoretic mechanism. The first attempt in this direction was carried out by Dolgov in \cite{Dolgov}. 

In this paper, we exploited the effective action of massless QED \cite{Basem2} to discuss this scenario. Although non-local actions defined over curved space are quite cumbersome, we showed how to cast the anomalous portion of the action into a usable form that resembles the starting Lagrangian for plenty of models that exist in the literature. In particular, we found the spectral index to depend on both the number of e-folds, the number of charged particles that run in the loop and most importantly on the sign of the beta function. With a positive beta function and dialing up the number of fermions we obtained a rather blue spectrum at the end of inflation. Demanding magnetic flux conservation, we found that a very low reheating temperature is required to produce a present day magnetic field consistent with the lower bound inferred on the IGM field \cite{Neronov,Taylor,Vovk}.

There is an important caveat about our presentation: the photon is not an active degree of freedom before spontaneous symmetry breaking. Thus one should ideally perform the analysis for the gauge bosons of the whole electroweak sector and evolve the system down to $T_{\text{EW}} = 100\, \text{GeV}$ before projecting onto the photon field. In this regard our analysis is exploratory. One important lesson is the effect of altering the sign of the beta function on the final result. In particular, it is possible to obtain a (nearly) scale-invariant spectrum with a negative beta function and an appropriate number of particles in the loop. An exciting future direction is to include gravitational loops in the presence of a positive cosmological constant. As emphasized by Toms in \cite{Toms}, the latter can render QED asymptotically free. We will hopefully pursue various directions and report on our findings in a future publication. 

\section{Acknowledgments}
I would like to thank Lorenzo Sorbo for plenty of useful discussions and for carefully reading the manuscript. Also, many thanks to John Donoghue for providing feedback on the manuscript. This work has been supported in part by the U.S. National Sciemce Foundation Grant No. PHY-1205896.



\begin{thebibliography}{99}


\bibitem{Brandenburg} 
  A.~Brandenburg and K.~Subramanian,
  ``Astrophysical magnetic fields and nonlinear dynamo theory,''
  Phys.\ Rept.\  {\bf 417}, 1 (2005).  
  
\bibitem{Neronov} 
  A.~Neronov and I.~Vovk,
  ``Evidence for strong extragalactic magnetic fields from Fermi observations of TeV blazars,''
  Science {\bf 328}, 73 (2010).  
  
\bibitem{Taylor} 
  A.~M.~Taylor, I.~Vovk and A.~Neronov,
  ``Extragalactic magnetic fields constraints from simultaneous GeV-TeV observations of blazars,''
  Astron.\ Astrophys.\  {\bf 529}, A144 (2011).  

\bibitem{Vovk} 
  I.~Vovk, A.~M.~Taylor, D.~Semikoz and A.~Neronov,
  ``Fermi/LAT observations of 1ES 0229+200: implications for extragalactic magnetic fields and background light,''
  Astrophys.\ J.\  {\bf 747}, L14 (2012). 
  
    
\bibitem{Durrer} 
  R.~Durrer and A.~Neronov,
  ``Cosmological Magnetic Fields: Their Generation, Evolution and Observation,''
  Astron.\ Astrophys.\ Rev.\  {\bf 21}, 62 (2013).
  
\bibitem{Turner} 
  M.~S.~Turner and L.~M.~Widrow,
  ``Inflation Produced, Large Scale Magnetic Fields,''
  Phys.\ Rev.\ D {\bf 37}, 2743 (1988).      
  
\bibitem{Demozzi} 
  V.~Demozzi, V.~Mukhanov and H.~Rubinstein,
  ``Magnetic fields from inflation?,''
  JCAP {\bf 0908}, 025 (2009).
 
\bibitem{Ratra} 
  B.~Ratra,
  ``Cosmological 'seed' magnetic field from inflation,''
  Astrophys.\ J.\  {\bf 391}, L1 (1992).
  
\bibitem{Martin} 
  J.~Martin and J.~Yokoyama,
  ``Generation of Large-Scale Magnetic Fields in Single-Field Inflation,''
  JCAP {\bf 0801}, 025 (2008).  
  
\bibitem{Carroll} 
  W.~D.~Garretson, G.~B.~Field and S.~M.~Carroll,
  ``Primordial magnetic fields from pseudoGoldstone bosons,''
  Phys.\ Rev.\ D {\bf 46}, 5346 (1992).
  
\bibitem{Anber} 
  M.~M.~Anber and L.~Sorbo,
  ``N-flationary magnetic fields,''
  JCAP {\bf 0610}, 018 (2006).  
  
\bibitem{Caprini} 
  C.~Caprini and L.~Sorbo,
  ``Adding helicity to inflationary magnetogenesis,''
  JCAP {\bf 1410}, no. 10, 056 (2014).     
    
\bibitem{Zakharov} 
  A.~D.~Dolgov and V.~I.~Zakharov,
  ``On Conservation of the axial current in massless electrodynamics,''
  Nucl.\ Phys.\ B {\bf 27}, 525 (1971).  
  
\bibitem{Frishman} 
  Y.~Frishman, A.~Schwimmer, T.~Banks and S.~Yankielowicz,
  ``The Axial Anomaly and the Bound State Spectrum in Confining Theories,''
  Nucl.\ Phys.\ B {\bf 177}, 157 (1981).
  
\bibitem{Horejsi2} 
  J.~Horejsi and M.~Schnabl,
  ``Dispersive derivation of the trace anomaly,''
  Z.\ Phys.\ C {\bf 76}, 561 (1997).
  
\bibitem{Basem3} 
  J.~F.~Donoghue and B.~K.~El-Menoufi,
  ``QED trace anomaly, non-local Lagrangians and quantum Equivalence Principle violations,''
  JHEP {\bf 1505}, 118 (2015).
  
\bibitem{Basem1} 
  J.~F.~Donoghue and B.~K.~El-Menoufi,
  ``Nonlocal quantum effects in cosmology: Quantum memory, nonlocal FLRW equations, and singularity avoidance,''
  Phys.\ Rev.\ D {\bf 89}, no. 10, 104062 (2014).
  
\bibitem{Espriu} 
  D.~Espriu, T.~Multamaki and E.~C.~Vagenas,
  ``Cosmological significance of one-loop effective gravity,''
  Phys.\ Lett.\ B {\bf 628}, 197 (2005).
  
\bibitem{Cabrer} 
  J.~A.~Cabrer and D.~Espriu,
  ``Secular effects on inflation from one-loop quantum gravity,''
  Phys.\ Lett.\ B {\bf 663}, 361 (2008).
  
\bibitem{Woodard1}
  R.~P.~Woodard,
  ``Perturbative Quantum Gravity Comes of Age,''
  Int.\ J.\ Mod.\ Phys.\ D {\bf 23} (2014) 09,  1430020.
  
\bibitem{Deser}
  S.~Deser and R.~P.~Woodard,
  ``Nonlocal Cosmology,''
  Phys.\ Rev.\ Lett.\  {\bf 99} (2007) 111301.              
  
\bibitem{Calmet} 
  X.~Calmet, D.~Croon and C.~Fritz,
  ``Non-locality in Quantum Field Theory due to General Relativity,''
  arXiv:1505.04517 [hep-th]. 
  
\bibitem{Conroy} 
  A.~Conroy, T.~Koivisto, A.~Mazumdar and A.~Teimouri,
  ``Generalized quadratic curvature, non-local infrared modifications of gravity and Newtonian potentials,''
  Class.\ Quant.\ Grav.\  {\bf 32}, no. 1, 015024 (2015).
  
\bibitem{Maggiore} 
  M.~Maggiore,
  ``Dark energy and dimensional transmutation in $R^2$ gravity,''
  arXiv:1506.06217 [hep-th].  
  
\bibitem{Basem2} 
  J.~F.~Donoghue and B.~K.~El-Menoufi,
  ``Covariant non-local action for massless QED and the curvature expansion,''
  JHEP {\bf 1510}, 044 (2015).     
  
\bibitem{Dolgov} 
  A.~Dolgov,
  ``Breaking of conformal invariance and electromagnetic field generation in the universe,''
  Phys.\ Rev.\ D {\bf 48}, 2499 (1993).

      
\bibitem{Agullo} 
  I.~Agullo and J.~Navarro-Salas,
  ``Conformal anomaly and primordial magnetic fields,''
  arXiv:1309.3435 [gr-qc].  
  
\bibitem{Kunze} 
  K.~E.~Kunze,
  ``Large scale magnetic fields from gravitationally coupled electrodynamics,''
  Phys.\ Rev.\ D {\bf 81}, 043526 (2010).           

\bibitem{Campanelli} 
  L.~Campanelli, P.~Cea, G.~L.~Fogli and L.~Tedesco,
  ``Inflation-Produced Magnetic Fields in $R^n F^2$ and $I F^2$ models,''
  Phys.\ Rev.\ D {\bf 77}, 123002 (2008).
  
\bibitem{Mazzitelli} 
  F.~D.~Mazzitelli and F.~M.~Spedalieri,
  ``Scalar electrodynamics and primordial magnetic fields,''
  Phys.\ Rev.\ D {\bf 52}, 6694 (1995).  
  
\bibitem{Prokopec} 
  T.~Prokopec, O.~Tornkvist and R.~P.~Woodard,
  ``Photon mass from inflation,''
  Phys.\ Rev.\ Lett.\  {\bf 89}, 101301 (2002).
  
\bibitem{Buchbinder} 
  I.~L.~Buchbinder, S.~D.~Odintsov and I.~L.~Shapiro,
  ``Effective action in quantum gravity,''
  Bristol, UK: IOP (1992) 413 p.  
 
\bibitem{Birrell}
  N.~D.~Birrell and P.~C.~W.~Davies,
  ``{\it Quantum Fields in Curved Space},''
(Cambridge University Press, Cambridge, 1982).

\bibitem{parkertoms}
L.~Parker and D.~Toms, "{\it Quantum Field Theory in Curved Spacetime, Quantum Fields and Gravity}"
(Cambridge University Press, Cambridge, 2009). 
  
\bibitem{Donoghue1} 
  J.~F.~Donoghue,
  ``General relativity as an effective field theory: The leading quantum corrections,''
  Phys.\ Rev.\ D {\bf 50}, 3874 (1994).
  
\bibitem{Donoghue2} 
  J.~F.~Donoghue,
  ``The effective field theory treatment of quantum gravity,''
  AIP Conf.\ Proc.\  {\bf 1483}, 73 (2012).
  
\bibitem{Isham} 
  S.~Deser, M.~J.~Duff and C.~J.~Isham,
  ``Nonlocal Conformal Anomalies,''
  Nucl.\ Phys.\ B {\bf 111}, 45 (1976).
  
\bibitem{Barvinsky1} 
  A.~O.~Barvinsky and G.~A.~Vilkovisky,
  ``The Generalized Schwinger-Dewitt Technique in Gauge Theories and Quantum Gravity,''
  Phys.\ Rept.\  {\bf 119}, 1 (1985).  
  
\bibitem{Barvinsky2} 
  A.~O.~Barvinsky and G.~A.~Vilkovisky,
  ``Covariant perturbation theory. 2: Second order in the curvature. General algorithms,''
  Nucl.\ Phys.\ B {\bf 333}, 471 (1990).
  
\bibitem{Barvinsky3} 
  A.~O.~Barvinsky and G.~A.~Vilkovisky,
  ``Covariant perturbation theory. 3: Spectral representations of the third order form-factors,''
  Nucl.\ Phys.\ B {\bf 333}, 512 (1990).

\bibitem{Barnaby} 
  N.~Barnaby and M.~Peloso,
  ``Large Nongaussianity in Axion Inflation,''
  Phys.\ Rev.\ Lett.\  {\bf 106}, 181301 (2011).

  
\bibitem{Namba} 
  N.~Barnaby, R.~Namba and M.~Peloso,
  ``Observable non-gaussianity from gauge field production in slow roll inflation, and a challenging connection with magnetogenesis,''
  Phys.\ Rev.\ D {\bf 85}, 123523 (2012).        
  
\bibitem{Dimo} 
  K.~Dimopoulos,
  ``Can a vector field be responsible for the curvature perturbation in the Universe?,''
  Phys.\ Rev.\ D {\bf 74}, 083502 (2006).  
        

\bibitem{Agashe} 
  K.~A.~Olive {\it et al.} [Particle Data Group Collaboration],
  ``Review of Particle Physics,''
  Chin.\ Phys.\ C {\bf 38}, 090001 (2014).  
  
\bibitem{Remmen} 
  G.~N.~Remmen and S.~M.~Carroll,
  ``How Many $e$-Folds Should We Expect from High-Scale Inflation?,''
  Phys.\ Rev.\ D {\bf 90}, no. 6, 063517 (2014).  
    
         
\bibitem{Jedamzik} 
  R.~Banerjee and K.~Jedamzik,
  ``The Evolution of cosmic magnetic fields: From the very early universe, to recombination, to the present,''
  Phys.\ Rev.\ D {\bf 70}, 123003 (2004).
  
\bibitem{Ringeval} 
  J.~Martin and C.~Ringeval,
  ``Inflation after WMAP3: Confronting the Slow-Roll and Exact Power Spectra to CMB Data,''
  JCAP {\bf 0608}, 009 (2006).
  
\bibitem{Peebles} 
  P.~J.~E.~Peebles and A.~Vilenkin,
  ``Quintessential inflation,''
  Phys.\ Rev.\ D {\bf 59}, 063505 (1999).    
  
\bibitem{Toms} 
  D.~J.~Toms,
  ``Cosmological constant and quantum gravitational corrections to the running fine structure constant,''
  Phys.\ Rev.\ Lett.\  {\bf 101}, 131301 (2008).            
          

\end{thebibliography}
\end{document}